# The Potential of $Ti_3C_2T_X$ Nano-sheets (MXenes) for Nanoscale Solid Lubrication Revealed by Friction Force Microscopy

A. Rodriguez,[1] M. S. Jaman,[1] O. Acikgoz,[1] B. Wang,[2] J. Yu,[2] P.G. Grützmacher,[3] A. Rosenkranz,[4,*] M.Z. Baykara[1,*]

[1] Department of Mechanical Engineering, University of California Merced, Merced, California 95343, USA

[2] Key Laboratory of Marine New Materials and Related Technology, Ningbo Institute of Material Technology & Engineering, Chinese Academy of Sciences, Ningbo 315201, People's Republic of China

[3] Institute of Engineering Design and Product Development, Tribology Research Group, TU Wien, 1060 Vienna, Austria

[4] Department of Chemical Engineering, Biotechnology and Materials, FCFM, Universidad de Chile, Santiago, Chile

* arosenkranz@ing.uchile.cl, mehmet.baykara@ucmerced.edu

## Abstract

$Ti_3C_2T_X$ nano-sheets (MXenes) are an emerging class of two-dimensional materials with outstanding potential to be employed in energy storage, catalysis, and triboelectric applications, on length scales ranging from the nano- to macroscopic. Despite rapidly accelerating interest in this new class of materials, their nanoscale frictional properties and in particular, their potential for solid lubrication on the nanoscale, have not been explored in detail yet. In this short communication, we present the first results on the nanoscale frictional characteristics of MXenes, in the form of a friction map obtained on an isolated $Ti_3C_2T_x$ nano-sheet deposited on a silicon dioxide substrate via friction force microscopy. Our experiments reveal that few-layer $Ti_3C_2T_x$ nano-sheets indeed act as solid lubricants on the nanoscale, reducing friction on the silicon dioxide substrates, although not as effectively as few-layer graphene. The results reported here pave the way for further studies focusing on nanoscale solid lubrication achieved by $Ti_3C_2T_X$ (MXene) nano-sheets.

## 1. Introduction

The discovery of MXene nano-sheets by Gogotsi's group in 2011 significantly expanded the family of 2D layered nano-materials [1–3]. In the initial study presented by Naguib *et al.* [2], the MAX-phase $Ti_3AlC_2$ was used as a precursor and chemically treated using a highly concentrated HF solution [1]. During this treatment, the aluminum layers of the original MAX-phase were selectively removed and replaced by -O, -OH and -F surface terminations thus creating $Ti_3C_2$ nano-sheets [3,4]. Due to the resulting surface terminations, the as-fabricated material was abbreviated as $Ti_3C_2T_x$, for which $T_x$ represents the variety of possible combinations of different surface terminations. After the first successful synthesis of $Ti_3C_2T_x$ nano-sheets, numerous research endeavors were conducted to develop milder, less risky etching routes, to increase the efficiency of the synthesis (processing times and synthesized amount) and to explore their outstanding properties in different fields such as energy storage, catalysis and water purification [3,5–9]. Moreover, various MAX phases based upon different early transition metals were utilized to synthesize molybdenum-, vanadium- and niobium-based MXene nano-sheets. Until today, the MXene family already consists of more than 30 experimentally-verified members with variations in the early transition metal, the stoichiometry as well as mixtures between carbidic and nitridic structures [3].

From a structural point of view, $Ti_3C_2T_x$ nano-sheets demonstrate great similarities with graphene nano-sheets due to their 2D layered-like structure with layer-by-layer distances of less than 1 nm. The 2D structure together with their chemical bonding and the resulting mechanical properties make these nano-sheets promising candidates in the field of tribology. This idea was supported by numerical studies using density functional theory (DFT) calculations and molecular dynamics (MD) simulations demonstrating low interlayer friction for $Ti_3C_2T_x$ nano-sheets and suggesting solid lubrication capabilities [10]. Encouraged by these numerical results, the tribological community started to explore the frictional performance of $Ti_3C_2T_x$ nano-sheets. Initially, these nano-sheets were used as lubricant additives [11–13] and reinforcement phases in composite materials [14–18]. In 2018/19, the first studies regarding the solid lubrication ability of $Ti_3C_2T_x$ nano-sheets were published [19–21]. Afterwards, the tribological performance of heterostructures combining $Ti_3C_2T_x$ nano-sheets with nano-diamonds or graphene quantum dots proved the outstanding tribological performance of these nano-sheets on the micro/macro-scale [22,23]. The wear behavior of these nano-sheets is particularly interesting due to the formation of ultra-wear resistant tribo-layers, which induce wear-free sliding conditions. However, a careful check of the existing literature evidently shows that nanoscale tribological research on $Ti_3C_2T_x$ nano-sheets was just initiated. Nanotribological properties of $Ti_3C_2T_x$ nano-sheets using atomic and/or friction force microscopy (AFM and/or FFM) have been barely reported, with only one study that does not focus on individual nano-sheets [24].

Motivated by the lack of nanotribological studies performed on $Ti_3C_2T_x$ nano-sheets in the literature and to put theoretical predictions of nanoscale lubricative properties [10] to test, we present here the first results of FFM experiments performed on isolated $Ti_3C_2T_x$ nano-sheets. Our measurements demonstrate that $Ti_3C_2T_x$ nano-sheets can indeed be employed as nanoscale solid lubricants, with results that are somewhat inferior to those achieved by well-known solid lubricants graphene and $MoS_2$.

## 2. Experimental details

### 2.1 Characterization of the as-synthesized $Ti_3C_2T_x$ nano-sheets

High-resolution transmission electron microscopy (HR-TEM) equipped with energy dispersive X-ray spectroscopy (TEM-EDX) using an EDAX detector was utilized to assess the overall quality and chemical composition of the as-synthesized $Ti_3C_2T_x$ nano-sheets. For imaging and chemical analysis, an acceleration voltage of 200 kV was utilized. In-situ diffuse reflectance infrared Fourier transform spectroscopy (DRIFTS) was performed to characterize the surface chemistry of the $Ti_3C_2T_x$ nano-sheets. For this purpose, a Nicolet iS50 infrared spectrometer (*Thermo Scientific*) with an MCT detector cooled by liquid nitrogen, a Ge-on-KBr On-Axis beam splitter and DiffusIR Environmental Chamber (*Pyke*) were used. In order to obtain spectra with a good signal-to-noise ratio, 200 scans with a resolution of 8 cm$^{-1}$ were collected. The sample of interest was directly placed into the sample holder without any pre-treatment (packing or dilution), flushed with Argon and heated up to the desired temperature. In order to analyse the intercalated water present between the nano-sheets, temperature-programmed decomposition mass-spectroscopy (TPD-MS) was performed on the nano-sheets without any pre-treatment using the *3Flex* from *Micromeritics* coupled with a mass spectrometer (*Cirrus 2, MKS Spectra Product*). For this purpose, the nano-sheets (15 - 30 mg) were heated in a quartz reactor (helium atmosphere with a flow rate of 100 mL/min) with a heating rate of 10 °C/min until a final temperature of 800 ºC was reached. Water formed during this treatment was characterized by mass spectrometry (m/z 18). The surface chemistry of the $Ti_3C_2T_x$ nano-sheets was investigated by X-Ray photoelectron spectroscopy (XPS, *Physical Electronics 1257*) using non-monochromatic MgKα radiation. For this analysis, the XPS system was operated at 15 kV and 400 W. All scans were acquired using a pass energy of 44.75 eV. Narrow scans of the $C_{1s}$, $F_{1s}$, $O_{1s}$ and $Ti_{2p}$ peaks were acquired using a step size of 0.1 eV. For the $C_{1s}$, $F_{1s}$ and $O_{1s}$ peaks, the Shirley method was used to subtract the background. In case of the $Ti_{2p}$ peak, a linear background subtraction was utilized. All photoelectron peaks were fitted using Lorentz-Gaussian functions.

## 2.2 Friction force microscopy measurements

Friction force microscopy (FFM) experiments were performed with a commercial atomic force microscope (*Bruker, Dimension 5000*) under ordinary laboratory conditions. $Ti_3C_2T_x$ nano-sheets in powder form were dispersed in a glass container by the addition of acetone, followed by ultrasonication for 10 minutes. The resulting solution was then deposited onto a $SiO_2$ substrate via drop-casting using a micropipette (droplet volume: 0.5 mL). The $SiO_2$ substrate was pre-heated up to 150 °C for ~5 minutes to facilitate acetone's evaporation and thus achieve a more homogenous dispersion of nano-sheets on the substrate. All FFM measurements presented here were conducted using two silicon cantilevers (Nanosensors, PPP-CONTR) with normal spring constant values of 0.61 N/m and 0.42 N/m, as determined by the Sader method [25]. During measurements, the applied normal load ranged from a few tenths of nN to tens of nN. Scanning was performed at rates of 1 to 2 Hz on areas of a few hundred nm and above in lateral size, whereby topography and lateral force maps were collected simultaneously. Friction values (in arbitrary units) were extracted from lateral force maps in the forward (trace) and backward (retrace) directions, based upon established procedures [26].

### 3. Results and discussion

### 3.1 Characterization of the as-synthesized $Ti_3C_2T_x$ nano-sheets

The HR-TEM micrograph (Figure 1(a)) reveals the multi-layered structure of the synthesized nano-sheets with an average layer distance of about 0.87 ± 0.07 nm and a total number of layers of about 10. TEM-EDX (Figure 1(b)) verifies titanium, carbon, fluorine and oxygen as main elements. Negligible traces of aluminium (0.3 wt.-%) were detected, which stem from the original MAX-phase $Ti_3AlC_2$. The obtained results by HR-TEM and TEM-EDX match well with previously published reports [27,28]. Figure 1(c) summarizes the results obtained by in-situ DRIFTS performed as a function of temperature. Irrespective of the temperature, pronounced signals at around 1176, 3360, 3380, and 3458 $cm^{-1}$ were detected. The first signal observed at 1176 $cm^{-1}$ can be attributed to C-F stretching in alkyl halides (1000 - 1400 $cm^{-1}$) [29], while the latter ones (3360, 3380, and 3458 $cm^{-1}$) imply the presence of -OH surface terminations. The existence of the surface terminations can be explained by the synthesis process. During etching, which involves HF for the chemical exfoliation of $Ti_3C_2T_x$ nano-sheets, the aluminium layers of the MAX-phase $Ti_3AlC_2$ are selectively etched out and randomly replaced by -O, -OH and -F surface terminations [3,30,31]. The TPD-MS used to characterize the water release from the nano-sheets shows three distinct peaks (Figure 1(d)). The first one located at 83 °C can be connected to superficial water. Moreover, the second peak and the tail observed for temperatures between 200 and 400 °C indicate the presence of adsorbed and strongly-bonded intercalated water, respectively [31–34]. Figure 1(e) and (f) show the results of the XPS analysis for the $Ti_{2p}$ and $C_{1s}$ peaks. As can be seen in Figure 1(e), three doublets were needed to fit the measured data for the $Ti_{2p}$ peak. In this context, the strongest contribution (62 %) located at 454.9 and 460.1 eV is associated with titanium carbides (MXenes) [31,35,36]. The contribution located at 456.1 and 461.4 eV, which accounts for about 17 %, can be connected to C-Ti-OH, which is in good agreement with results of in-situ DRIFTS [31]. The third contribution accounting for 21%, which is located at 457.8 and 462.7 eV, can be associated with C-Ti-O [31,37]. The peak fitting of the $C_{1s}$ photoelectron peak (Figure 1(f)) required four contributions. The most pronounced contribution is located at 284.8 eV, which represents carbon in C-C chemical environment and adventitious carbon [38]. The contribution at 281.8 eV is assigned to C-Ti-

$T_x$ [37,38]. Organic compounds (hydrocarbons) are well represented in the spectrum by the contribution located at 286.5 eV [35,37]. A less pronounced signal was found for C-F states at binding energies of 288.5 eV [38]. In summary, the XPS analysis confirms -O, -OH and -F surface terminations as discussed for the results of in-situ DRIFTS.

### 3.2 Friction force microscopy measurements

In order to evaluate the potential of $Ti_3C_2T_x$ nano-sheets as solid lubricants on the nanoscale, FFM measurements were conducted. While a large portion of the $SiO_2$ substrate surface was covered by a film consisting of densely packed $Ti_3C_2T_x$ nano-sheets after drop-casting, regions on the substrate comprising isolated nano-sheets were occasionally observed, which allowed for the evaluation of their nanoscale lubricative properties. Figure 2(a) shows a friction map recorded on such a region, where an isolated $Ti_3C_2T_x$ nano-sheet with a triangular morphology and a height of about ~6.5 nm (corresponding to less than 8 layers) is located. While the friction map (recorded with an applied normal load of 77 nN) directly confirms a reduction of friction on the $Ti_3C_2T_x$ nano-sheet when compared with the $SiO_2$ substrate, the line profile extracted from the friction map (Figure 2(b)) allows for a quantitative evaluation. In particular, we observe that the friction recorded on the $Ti_3C_2T_x$ nano-sheet is ~35% of that recorded on the $SiO_2$ substrate. This level of solid lubrication performance, while remarkable, is somewhat inferior to that verified for $MoS_2$ on $SiO_2$, for which reductions in the friction force down to 15-40% were reported [39,40], and for graphene on $SiO_2$, for which reductions in friction force down to 10-15% were presented [40–42].

To further explore the nanotribological properties of $Ti_3C_2T_x$ nano-sheets, we have performed load-dependent friction measurements on an area containing multiple nano-sheets, whereby the applied load was varied from 0.7 nN to 26.3 nN (Figure 3). The results demonstrate that friction on $Ti_3C_2T_x$ nano-sheets monotonically increases with applied normal load, in accordance with the majority of previous nanotribological studies in the literature performed via FFM [43].

While the results for the load dependence of friction reported in Figure 3 are perhaps not surprising, it is remarkable that the $Ti_3C_2T_x$ nano-sheets exhibit nanoscale solid lubrication properties, much like well-known solid lubricants graphene and $MoS_2$. Solid lubrication exhibited by such lamellar materials is often attributed to weak van-der-Waals forces acting between layers [44,45]. Zhang *et al.* predicted via computational approaches that the energy barrier for inter-layer sliding between $Ti_3C_2T_x$ nano-sheets is in the same range as that of $MoS_2$ [10], indicative of comparable solid lubrication characteristics. On the other hand, Hu *et al.* calculated that the interlayer coupling between $Ti_3C_2T_x$ nano-sheets, although weakened by surface groups such as $-O_2$, are still 2 to 4 times stronger than that for graphene and $MoS_2$ [46]. These results were partially supported by Zhang *et al.,* who predicted $Ti_3C_2T_x$ nano-sheets to be good solid lubricants despite exhibiting significantly higher energy barriers to sliding than graphene [47]. These predictions are now confirmed by our FFM experiments, which verify notable solid lubrication properties for $Ti_3C_2T_x$ nano-sheets that are somewhat inferior to those exhibited by graphene and $MoS_2$.

Assuming that no transfer of $Ti_3C_2T_x$ nano-sheets to the tip apex takes place during the experiments, it may be argued that interlayer sliding between $Ti_3C_2T_x$ nano-sheets does not play an essential role in FFM measurements. If that is indeed the case, other mechanisms leading to low friction at the tip-$Ti_3C_2T_x$ interface, including but not limited to those involving topographical roughness and/or suppressed chemical interactions of the tip apex with the surface groups on the $Ti_3C_2T_x$ nano-sheets would need to be explored. Last but not least, the role that intercalated water molecules (as verified by the results of TPD-MS (Figure 1(d)) play in modulating phonon-based energy dissipation mechanisms, such as those studied in FFM-based nanotribological studies of graphene [48,49], also constitutes an important avenue of future investigation.

## 4. Conclusions

We presented here the first experimental results on the nanoscale solid lubrication characteristics of $Ti_3C_2T_X$ nano-sheets obtained via FFM. Friction maps revealed that few-layer $Ti_3C_2T_X$ nano-sheets indeed act as solid lubricants on $SiO_2$ substrates, with friction mitigation capabilities that are somewhat inferior to that of graphene and $MoS_2$. Further work needs to be performed in order to elucidate the mechanisms behind the lubricative character of $Ti_3C_2T_X$ nano-sheets and evaluate their effect on relevant applications such as triboelectric nanogenerators [50].

**Acknowledgements**

This work was supported by the Merced Nanomaterials Center for Energy and Sensing (MACES) via the National Aeronautics and Space Administration (NASA) [Grant No. NNX15AQ01]. A. Rosenkranz gratefully acknowledges the financial support given by ANID-CHILE within the project Fondecyt 11180121 and the VID of the University of Chile in the framework of "U-Inicia UI013/2018". A. Rosenkranz and B. Wang acknowledge the financial support of Chinese Academy of Sciences President's International Fellowship Initiative (2020VEC0006). V.M. Fuenzalida, F. Gracia, S. Suarez and N. Escalona are acknowledged for their valuable input regarding the characterization of the as-synthesized nano-sheets.

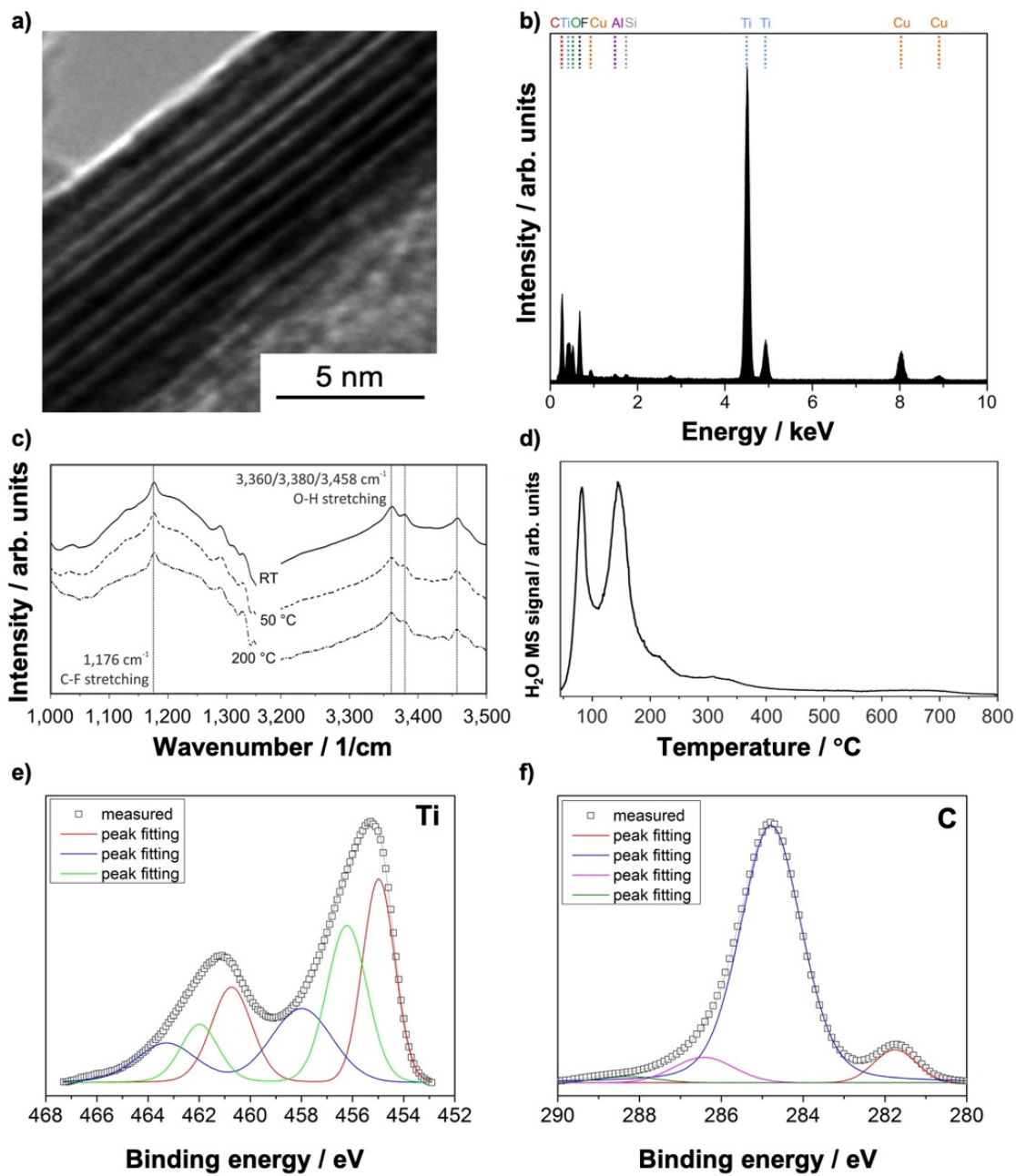

**Figure 1. a**) HR-TEM micrograph of the as-synthesized $Ti_3C_2T_x$ nano-sheets and (**b**) the corresponding TEM-EDX analysis. (**c**) summarizes the temperature-dependent DRIFTS results and (**d**) depicts the results of TDP-MS. (**e**) and (**f**) reveal the XPS results of the narrow scans acquired for the $C_{1s}$ and $Ti_{2p}$ regions with the measured data and the respective peak fittings. Please note that no correction due to surface charging has been applied and all peaks have been fitted using Lorentz-Gaussian functions.

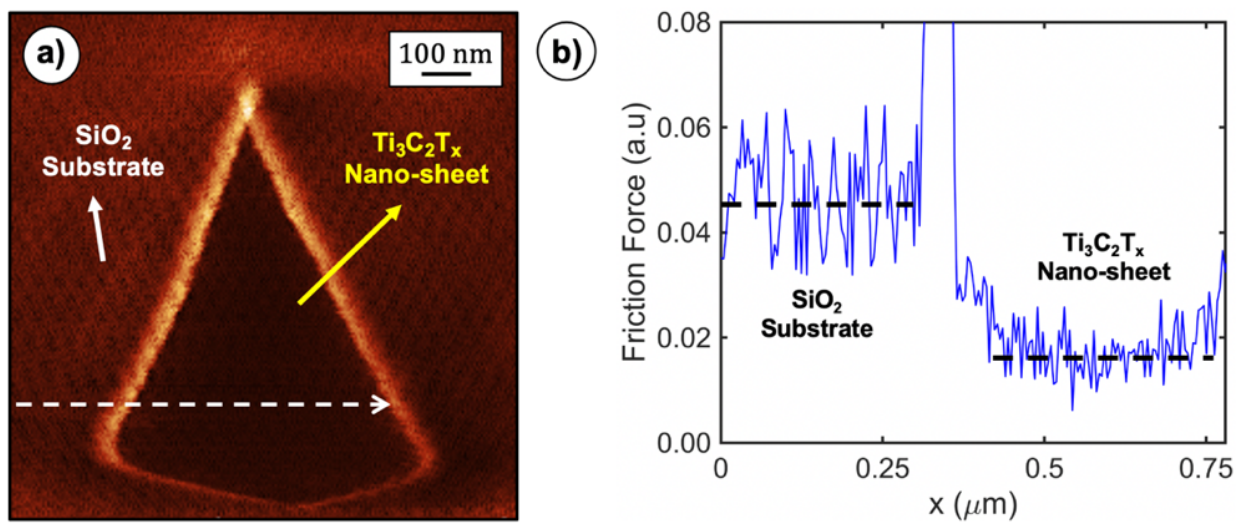

**Figure 2. (a)** Friction map recorded on an isolated, few-layer $Ti_3C_2T_x$ nano-sheet situated on the $SiO_2$ substrate. Please note that darker colors correspond to lower friction. **(b)** Line profile extracted from the friction map of **(a)** along the dashed white arrow, quantitatively demonstrating the observed friction reduction on the few-layer $Ti_3C_2T_x$ nano-sheet compared to the $SiO_2$ substrate.

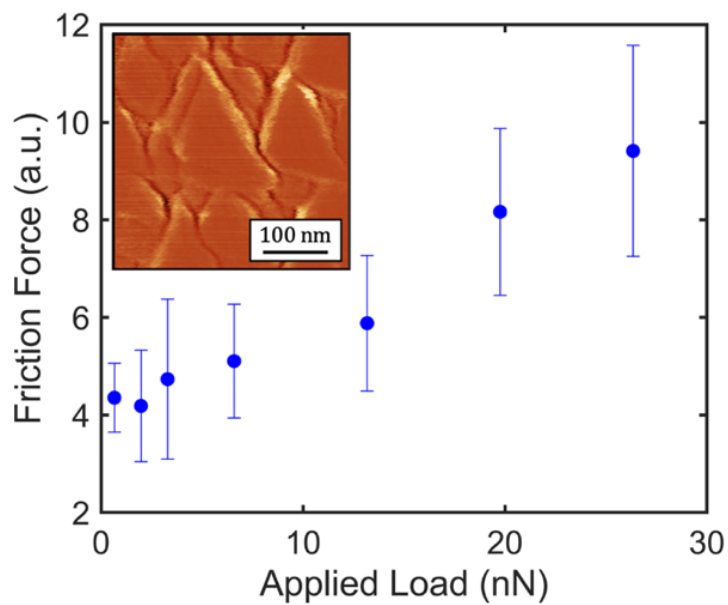

**Figure 3.** Mean friction as a function of the applied load measured on a film comprising multiple $Ti_3C_2T_x$ nano-sheets. Error bars represent the standard deviation extracted by dividing the corresponding friction map (inset) into a total of ten regions.